\documentclass[12pt,a4paper]{article}
\usepackage{amsmath}
\usepackage{amssymb}
\usepackage{overcite}
\usepackage{xfrac}
\usepackage{xcolor}

\title{\bf Molecular Similarity in Machine Learning of Energies in Chemical Reaction Networks}

\date{April 25, 2025}

\author{\vspace{0.3cm}
Stefan Gugler%
\footnote{ORCID: 0000-0001-6257-1923}
~and~Markus Reiher%
\footnote{Corresponding author; mreiher@ethz.ch; ORCID: 0000-0002-9508-1565}\\
\textit{Department of Chemistry and Applied Biosciences, ETH Z\"urich,}\\ 
\textit{Vladimir-Prelog-Weg 2, 8093 Z\"urich, Switzerland}\\[1ex]
}

\begin{document}

\maketitle

\renewcommand*{\thefootnote}{\fnsymbol{footnote}}

\vspace*{0.2cm}

\begin{abstract}
Machine learning has emerged as a powerful tool for predicting molecular properties in chemical reaction networks with reduced computational cost. However, accurately predicting energies of transition state (TS) structures remains a challenge due to their distinct electronic characteristics compared to stable intermediates. In this work, we investigate the limitations of structural descriptors in capturing electronic differences between minima and TS structures. We explore $\Delta$-machine learning approaches to predict correlation energy corrections using both Hartree--Fock (HF) and density functional theory (DFT) as reference methods. Our results demonstrate that learning the energy difference between DFT and coupled cluster methods outperforms direct learning and HF-based $\Delta$-learning. We also assess the effectiveness of combining electronic descriptors with structural ones but find that simple electronic features do not significantly enhance the prediction of TS energies. These findings highlight the need for more sophisticated descriptors or integrated approaches to accurately predict the electronic energies of TS structures within chemical reaction networks.
\end{abstract}

\section{Introduction}
\label{sec:intro}

Chemical reaction networks (CRNs) describe chemical processes and the reactivity of stable intermediates.
\cite{broadbelt1994, broadbelt1996, broadbelt2005, fialkowski2005, gothard2012, kowalik2012, sameera2016, dewyer2018, maeda2011, feinberg2019, simm2019, unsleber2020, baiardi2022}
They are a key concept for understanding
catalysis\cite{deutschmann1998,zhu2005,gossler2019,ulissi2017,steiner2022},
combustion\cite{sankaran2007,harper2011},
polymerization\cite{vinu2012},
and atmospheric chemistry\cite{vereecken2015}.
At the heart of CRN analysis is the computation of properties and reactivity
of the network's molecular structures and transition states (TSs).
These structures are stationary points (i.e., local minima and first-order saddle points) on the Born--Oppenheimer potential energy surface (PES),
which can be accessed by solving the electronic Schr\"odinger equation
for varying nuclear 
coordinates.\cite{jensen2017,cramer2004}
Thermodynamic contributions of the nuclear framework supplement these electronic energies to yield free energy data for microkinetic modeling.\cite{proppe2017,proppe2019b, Suleimanov2015,gao2016a,susnow1997,han2017}

Electronic structure models assign energies to any given set of coordinates,
which correspond to points on the Born--Oppenheimer hypersurface.
However, all these models suffer from
a fundamental trade-off between computational cost and accuracy,
as they scale with some power of system size.
This trade-off is a severe limitation for the exploration
of CRNs that involve large numbers of chemical structures.
\cite{kowalik2012,bajczyk2018,simm2019,jacob2018,margraf2019,steiner2022}
As a consequence,
less accurate but fast methods such as
semiempirical methods
are employed to accelerate the exploration.\cite{savoie2014,sameera2016,dewyer2018,jensen2019,unsleber2022}
This poses the risk that not only energies and structures of stable intermediates and their connecting first-order saddle points might be severely compromised,
but also the whole network topology.

A CRN produces a data structure where neighboring local minimum structures on the PES
are related to one another by elementary steps via TS structures.
\cite{jordan1979,laidler1984}
An elementary step is the smallest unit of change in a reaction.
Hence, any two minimum structures which are directly connected to each other by an elementary step
are structurally more related than an arbitrary set of molecular structures. Connecting TSs are then more similar to an end point of an elementary step than the two end points are with respect to one another.

Similarity-exploiting machine learning (ML) methods
\cite{scholkopf2018,hofmann2008,herbrich2001,vapnik1998}
such as Gaussian process (GP) regression\cite{rasmussen2006,deringer2021a,griffiths2022,mackay2005}
can potentially improve on limited accuracy at low computational cost.\cite{simm2018,reiher2022}
In a supervised learning setting,
where both input and target of the desired properties are known for the data set,\cite{bishop2006,mackay2005,murphy2012}
a certain number of highly accurate yet expensive reference calculations is required to train the ML model.
\cite{butler2018,himanen2019,batra2021,ramprasad2017,schmidt2019,schleder2019,dral2020,carleo2019}
In the worst case,
reference calculations would be required for
every molecular system occurring in the network.
Uncertainty quantification, a baked-in feature of GP regression,
is key for sampling new candidate structures
that require a reference calculation.\cite{simm2016,proppe2017,proppe2019a,simm2018,reiher2022}
In the case of high uncertainty,
additional reference calculations are to be launched on-the-fly
to further improve on the training data set
in a computing-time-economic fashion,
a procedure which is called active learning
\cite{fedorov1972,lewis1994,hastie2001,bishop2006,settles2010}
and has been used in chemistry in various applications (see Refs.
\cite{artrith2012,smith2018a,proppe2019a,zhang2019,jablonka2021} for examples).
GP regression is especially suitable in the small and intermediate data regime
of hundreds of data points,\cite{zhang2018}
rather than tens of thousands,
where a neural network might become more appropriate.\cite{aggarwal2018,ryu2019,zhang2019,hwang2020,scalia2020,gugler2020,eissler2025}

A key ingredient for the learning process is the so-called molecular descriptor which is
a computer readable representation of a molecule
that can be used to assess their similarity.
The related concept of a similarity measure (kernel\cite{mercer1909}) is a mathematical metric,
i.e., a function that measures the distance between two points in descriptor space.
The closer the two points are according to the metric, the more similar they are.
Although there are many descriptors and measures of similarity available,\cite{todeschini2009,jones2023}
here we focus on two established ones that we have analyzed in some detail in previous work:\cite{gugler2022}
the smooth overlap of atomic positions (SOAP)\cite{bartok2013}
and the eigenvalues of the Coulomb matrix (CM).\cite{rupp2012}
These measures are readily available in ML software designed for chemical applications. 
\cite{himanen2020,dral2019,collins2018,haghighatlari2019,khatib2020,musil2021,caro2019}
Furthermore, we developed two variants of those representations,
namely the smooth overlap of electron densities (SOED),
which we will also consider in this work,
and the Coulomb list (CL).\cite{gugler2022}
The SOAP family of representations aims at describing a local environment and is sometimes
called a local descriptor.\cite{de2016}
As we use them for entire molecules, we will review how to make them globally applicable.

In the case of a CRN, where all connected structures are similar, we arrive at a somewhat paradoxical situation:
while reactant and product molecules of neighboring vertices are related by construction (as they are connected by only one elementary reaction step),
the connecting TS structure is usually even more similar to either product or reactant. 
Yet, it is well known in electronic structure theory that TS structures,
if they represent activated molecules with one or two stretched chemical bonds,
present a very different electron correlation problem compared to stable reactant or product structures.
\cite{golubeva2007,kedziora2016,stein2017,manna2019,duan2020,vitillo2022}
Hence, this is a situation in which structural similarity seems to be
insufficient to simultaneously evaluate electronic similarity, which governs the molecular properties.
As a result, a molecular similarity measure might rightfully determine a TS structure to be more similar to
either side of the reaction arrow but at the same time miss the fact that its electronic structure will be rather
different.

We study the effectiveness of using both Hartree--Fock and density functional theory (DFT) as base methods and assess whether incorporating electronic descriptors alongside structural ones can improve the prediction of TS energies. Furthermore, we analyze different variations of SOAP and Coulomb interaction based descriptors and their effectiveness for prediction of minima and TS energies.

This work is organized as follows: 
In Section~\ref{sec:theory}, we present the theoretical background, including general considerations about ML approaches for predicting electronic energies,
SOAP and SOED as molecular similarity descriptors, and Coulomb interaction descriptors like the CM and CL. 
Section~\ref{sec:comp_method} outlines the computational methodology employed to generate our data sets and perform the electronic structure calculations. 
In Section~\ref{sec:results}, we present and discuss our results,
analyzing the elementary step similarity dilemma,
comparing $\Delta$-ML with direct learning,
examining technical considerations for SOAP descriptors,
exploring the prediction of TS energies from minima,
and investigating the combination of electronic information with structural descriptors.
Finally, we conclude in Section~\ref{sec:conclusion} with a summary of our findings.

\section{Theory}
\label{sec:theory}

\subsection{General Considerations}
Instead of learning the target energy directly (direct learning),
we aim at learning the difference between two models ($\Delta$-ML).\cite{ramakrishnan2015}
$\Delta$-ML is very popular in computational chemistry\cite{ramakrishnan2015,ruth2022,bogojeski2020,qu2022,mcdonagh2018}
because it can reduce the error on an expensive target quantity a lot with the help
of a cheaper, physical model that is usually readily at hand.
The expectation is that some of the complexity can be reduced for the ML model
by ``removing'' parts of the physics that can be efficiently captured with a low-cost physical model.\cite{bogojeski2020,stohr2020,dral2020,atz2021,ruth2022}
In our case, the target quantity is the error of a mean-field
description measured in terms of the correlation energy.\cite{lowdin1955}
Evaluating the electron correlation energy with a post-HF method requires substantial computational effort.
It is typically accurately approximated as the difference between 
    HF energies and
    coupled cluster (CC)
    including single and double excitations
    with the perturbative triples correction (CCSD(T))
    energies.  
The simplest post-HF method,
second order perturbation theory (MP2),\cite{moller1934}
has a $\mathcal{O}(N^5)$ computational complexity with $N$ being the number of basis functions.
The learnable difference to the HF energy provides the MP2 correlation energy.\cite{han2021a}
CC including singles and doubles excitations (CCSD) scales as $\mathcal{O}(N^6)$ and its corresponding
correlation energy has been learned exploiting MP2 amplitudes.\cite{margraf2019,mcdonagh2018}
CCSD(T), the method of our choice here, scales as $\mathcal{O}(N^7)$.
We add that the post-Hartree-Fock correlation energy was also predicted utilizing the Fock, Coulomb,
and exchange matrices obtained from HF calculations\cite{welborn2018,cheng2019}
or electron density and correlation energy density.\cite{morawietz2013,brockherde2017,nudejima2019,grisafi2019,unke2019,pederson2022}

We also study a difference with a less canonical interpretation than the electron correlation energy, which is essentially a measure of the deficiencies of HF rather than a difference that would lead from a trusted low-level model to CCSD(T).
Hence, we also consider the difference between DFT 
and CCSD(T).
In \citen{nandi2021} and \citen{bogojeski2020}, only a single PES was considered for MD.
For instance, in the former small molecules such as H$_3$O$^+$ were studied and spectroscopic accuracy ($<1$ cm$^{-1}$) was achieved.
In the DFT different to CC
we note that double counting effects could deteriorate
the learning, which requires careful investigation.
Tentatively, however, a more accurate low-accuracy base method
should improve the $\Delta$-ML model.

\subsection{Smooth overlap of atomic positions and of electron densities}
\label{sec:soap-ext}

We briefly review the SOAP formalism, which is key for assessing molecular similarity in this work.
SOAP describes a local, typically atom-centered environment
that remains invariant under translation, rotation, and permutation.\cite{bartok2013}
For the comparison of entire molecules,
the local formalism can be expanded to incorporate a global descriptor.\cite{de2016}
Previously, we employed the local descriptor at the center of mass as a relative location in the molecule to compare to other molecules.\cite{gugler2022}
In order to transition from a local to a global descriptor, 
all possible pairwise combinations of atomic environments (instead of only the center of mass) between two molecules are considered.\cite{de2016}
A straightforward but coarse-grained way to treat all possible pairwise combinations is to take their average.\cite{de2016}

The $I$th atom in a molecular structure $A$ at position $\mathbf{R}_I^{A}$ possesses an environment $\mathcal{X}_I^{A}$.
The similarity between two arbitrary environments in structures $A$ and $B$,
denoted as $\mathcal{X} \equiv \mathcal{X}_I^{A}$ and $\mathcal{X}' \equiv \mathcal{X}_J^{B}$,  is measured by

\begin{equation}
k\left(\mathcal{X}, \mathcal{X}^{\prime}\right)
=
\frac{
    \tilde{k}\left(\mathcal{X}, \mathcal{X}^{\prime}\right)
}{
    \sqrt{
        \tilde{k}(\mathcal{X}, \mathcal{X}) 
        \tilde{k}(\mathcal{X}^{\prime}, \mathcal{X}^{\prime})
    }
}\, ,
\end{equation}

where the denominator represents a normalization factor and we denote the unnormalized similarity measure by a tilde.
This similarity is related
to the Carb\'{o} index\cite{carbo1980} to measure the similarity based on the overlap of electron density.\cite{gugler2022}

We briefly derive $\tilde{k}$ following Ref. \citen{de2016} to introduce the formalism necessary for later elaborations in this work.
The SOAP density for an environment, $\mathcal{X}$, is defined as a sum of Gaussians
where each Gaussian is placed on an aufpunkt\cite{gugler2022} (that is, the coordinates of the Gaussian's center; usually those of an atomic nucleus)
inside the environment at position $\mathbf{R}_I$ and has a width $\sigma^2_I$,
sometimes expressed as reciprocal $\alpha_I = 1/(2\sigma_I^2)$, 

\begin{equation}
\rho
(\mathbf{r}; \{\mathbf{R}_I\}, \{\alpha_I\})
=\sum_{I=1}^{N}
    \exp
    \left(
        -\alpha_I
        \left| \mathbf{r} - \mathbf{R}_I\right|^2
        \right)
\label{eqn:soap}
\end{equation}

where $N$ is the number of aufpunkte in the environment and
it is usually assumed that $\alpha \leftarrow \alpha_I$,
so that all widths are the same.

It is possible to calculate an analytical overlap
over all possible rotations of two SOAP environment densities\cite{bartok2013,gugler2022},

\begin{equation}
\tilde{k}\left(\mathcal{X}, \mathcal{X}^{\prime}\right)
=\int
    \left|\int \rho(\mathbf{r}) \rho'(\hat{R} \mathbf{r}) \mathrm{d} \mathbf{r}\right|^n
\mathrm{d}\hat{R}
\, ,
\label{eqn:integral}
\end{equation}

where we omitted the parametric dependence on the coordinates, $\{\mathbf{R}_I\}$, and on the widths, $\{\alpha_I\}$.
The inner integral over the product of the two densities determines their overlap.
The relative orientation between the two densities is denoted by
by the rotation $\hat{R}$, specified by Euler angles.
The outer integral with the Haar measure\cite{nachbin1965} $\mathrm{d}\hat{R}$
integrates over all possible rotations in SO(3).\cite{bartok2013,gugler2022}

In most applications, $n$ is set to $2$.
This exponent allows one to generate the power spectrum, that is
the Fourier transform of the second-order cumulant of the densities.\cite{bartok2013,de2016}
The Fourier transform of the third-order cumulant-generating function,
i.e. $n=3$,
is called the bispectrum\cite{mendel1991}
and enables one to obtain the bispectrum-based SOAP kernel.\cite{bartok2013,musil2021}
This last variation
allows the kernel to distinguish between enantiomers,
which is not possible with standard SOAP with $n=2$.
Such an inability to distinguish between enantiomers is inherent to any two-body expansion ($n<3$) and
will increase the theoretically achievable smallest error,
which is often negligible\cite{pozdnyakov2020}
except along certain manifolds of degenerate structures.\cite{pozdnyakov2020,pozdnyakov2022,parsaeifard2022}

The scaffold density 
of Eq.\ (\ref{eqn:soap}) can be expressed in terms of
expansion coefficients, $c_{blm}$,
spherical harmonics, $Y_{lm}$,
and the corresponding orthonormal basis functions, $g_b$,
\begin{equation}
\rho(\mathbf{r}; \{\mathbf{R}_I\}, \{\alpha_I\})=
\sum_{b l m} c_{b l m} g_b(r) Y_{l m}(\hat{\mathbf{r}})
\label{eqn:expansion}
\end{equation}
where $\hat{\mathbf{r}}$ is the unit vector in the direction of $\mathbf{r}$.
The exponents $\alpha_I$ affect the expansion coefficients, $c_{blm}$.
The orthonormality of the basis functions implies $\int g_b(r) g_{b'}(r) dr = \delta_{bb'}$, where the Kronecker delta-function is defined as
$\delta_{bb'}=1$ if $b=b'$, and $0$ otherwise.
In the original publication on SOAP,\cite{bartok2013} 
the orthonormal basis functions were chosen to be polynomials
but primitive Gaussian-type orbitals can be used as well
(which would have an additional index, $l$, representing
the degree of the spherical harmonic.)\cite{himanen2020}

For two structures, $A$ and $B$, we can now create a matrix with all pairwise environments,
$k\left(\mathcal{X}_I^{A}, \mathcal{X}_J^{B}\right)$.
This matrix over $I$ and $J$ now contains all pairwise similarities between environments centered on every aufpunkt in both molecules,
turning it into a global descriptor.
In the center of mass definition\cite{gugler2022},
$I$ and $J$ point to the position of the centers of mass of $A$ and $B$, respectively.
Different ways\cite{de2016} have been proposed to obtain a scalar to quantify the similarity between two molecules.
One of the simplest is the average kernel, $\bar{K}(A,B)$, taking the average over all elements,
\begin{equation}
\bar{K}(A,B)
=\frac{1}{N_A N_B}
\sum_{I}^{N_A}
\sum_{J}^{N_B}
    k\left(\mathcal{X}_I^{A}, \mathcal{X}_J^{B}\right)
\label{eqn:average}
\end{equation}
for $N_A$ aufpuntke of molecule A and $N_B$ aufpunkte for molecule B.

While we expect that including all pairwise environments will increase the sensitivity to structural differences compared to the center of mass definition,
some sensitivity is lost due to averaging.

\subsection{Coulomb interaction descriptors}

The CM is defined as\cite{rupp2012}
\begin{equation}
C_{I J}=
\left\{\begin{array}{ll}
\displaystyle
\frac{Z_I^{2.4}}{2} & I=J \\
\displaystyle
\frac{Z_I Z_J}{\left|\mathbf{R}_I-\mathbf{R}_J\right|} & I \neq J
\end{array}\right. \, .
\end{equation}
The CM is often diagonalized to obtain the eigenvalue spectrum and thereby loses its physical interpretability. Therefore, we considered CL\cite{gugler2022}, which is related to the CM, but
neglects the diagonal and
sums over nuclear pair interactions of a given nucleus $J$,

$$
L_{J}=\sum_{
    \genfrac{}{}{0pt}{}{I}{I \neq J} 
} C_{I J} \ .
$$

A convenient feature of CL is that it correlates with the external potential and therefore it can be related to a fundamental quantity of electronic structure theory.\cite{gugler2022}
Hence, CL can be considered a suitable descriptor that relates to the underlying physics,
yet avoids any computationally costly components.

\section{Computational Methodology}
\label{sec:comp_method}

To generate a CRN that can serve as a testing ground for our analysis,
we employed the \textsc{Chemoton} software
\cite{unsleber2022} with Readuct\cite{brunken2021} and Puffin\cite{bensberg2023},
starting from $N$-acetylimidazole, C$_5$H$_6$N$_2$O.
We constrained the exploration to remain on the same PES
to avoid descriptor extensivity problems.\cite{gugler2022}
The exploration uses the semi-empirical tight-binding method
GFN2-xTB\cite{bannwarth2019}
to generate elementary steps and
their reactants, TS structures, and products.
We subsequently optimized the generated structures
with the PBE\cite{perdew1996a} density functional and the def2-SVP basis set\cite{weigend2006}
with the quantum chemistry software Orca 4.2.1\cite{neese2012,neese2018}.
The truncation of the exploration was determined at
401 structures
for which reference calculations with
CCSD(T)\cite{bartlett1990,raghavachari1989}
and the aug-cc-pVTZ\cite{dunning1989} basis sets were carried out with Orca.
For all calculations,
the keyword \texttt{TightSCF} was used together with a Hueckel guess.
Furthermore,
all calculations were carried out in an unrestricted formalism and without invoking a frozen-core approximation.
For a detailed description of how the reactions were selected and for all Cartesian coordinates of the structures, see the Supporting Information.

The ML method employed was GP regression\cite{rasmussen2006},
as implemented in the scikit-learn package.\cite{pedregosa2012}

In addition to the CRN data set created for this work,
we adopt a data set generated by Grambow et al.\cite{grambow2020}
These authors explored\cite{dewyer2018,grambow2018} a PES with the growing string method\cite{zimmerman2015},
yielding 12,000 reactions
with the $\omega$B97X-D3/def2-TZVP combination of exchange-correlation density functional and basis set.\cite{grambow2020}
These reactions are all unimolecular
gas-phase reactions with up to seven carbon, oxygen, or nitrogen atoms
(subsampled from GDB-17).\cite{ruddigkeit2012}).

\section{Results}
\label{sec:results}

\subsection{Elementary-step similarity dilemma}

The elementary-step similarity dilemma\cite{gugler2022}
refers to the observation that
the TS structure of a reaction is often more 
similar to the reactants or products than those are similar to each other.
However, with respect to their electronic structures,
TS structures can often be more distinct from minimum structures (such as reactants or products) because of a potential 
multi-configurational character emerging in homolytic bond stretches.

We note that it is the context of CRNs that allows for this dilemma to play out. In general, all structures within a CRN are clearly related and likely to be rather similar. By contrast, in a general ML attempt toward activation barriers
\cite{lewis-atwell2023,grambow2020c,heid2022,spiekermann2022,spiekermann2022a,choi2018,zhao2023,ismail2022, vangerwen2022, vangerwen2024, vangerwen2024a}
one would demand predictability across large parts of reaction space.
Typical of many approaches toward reaction variables is to model reactants and products in a single descriptor, for instance, by concatenating them or constructing a difference of the feature vectors.\cite{choi2018,vangerwen2024}
Employing a three-dimensional representation instead of a two-dimensional one (such as a SMILES string) has been shown to be beneficial for accurate modeling.\cite{vangerwen2024}
Apart from inspiration from cheminformatics-based descriptors\cite{heid2022} or traditional ML methods\cite{lewis-atwell2023,farrar2022},
many current approaches to reaction barrier prediction employ neural networks as large data sets have become available.\cite{choi2018,grambow2020c,spiekermann2022,spiekermann2022a,zhao2023,ismail2022}
Our work is different from such general attempts in that we restrict ourselves to local comparisons within one CRN and focus on accuracy of the prediction in relation to the molecular-similarity descriptor employed.

\begin{figure}
    \centering
    \includegraphics[width=0.9\textwidth]{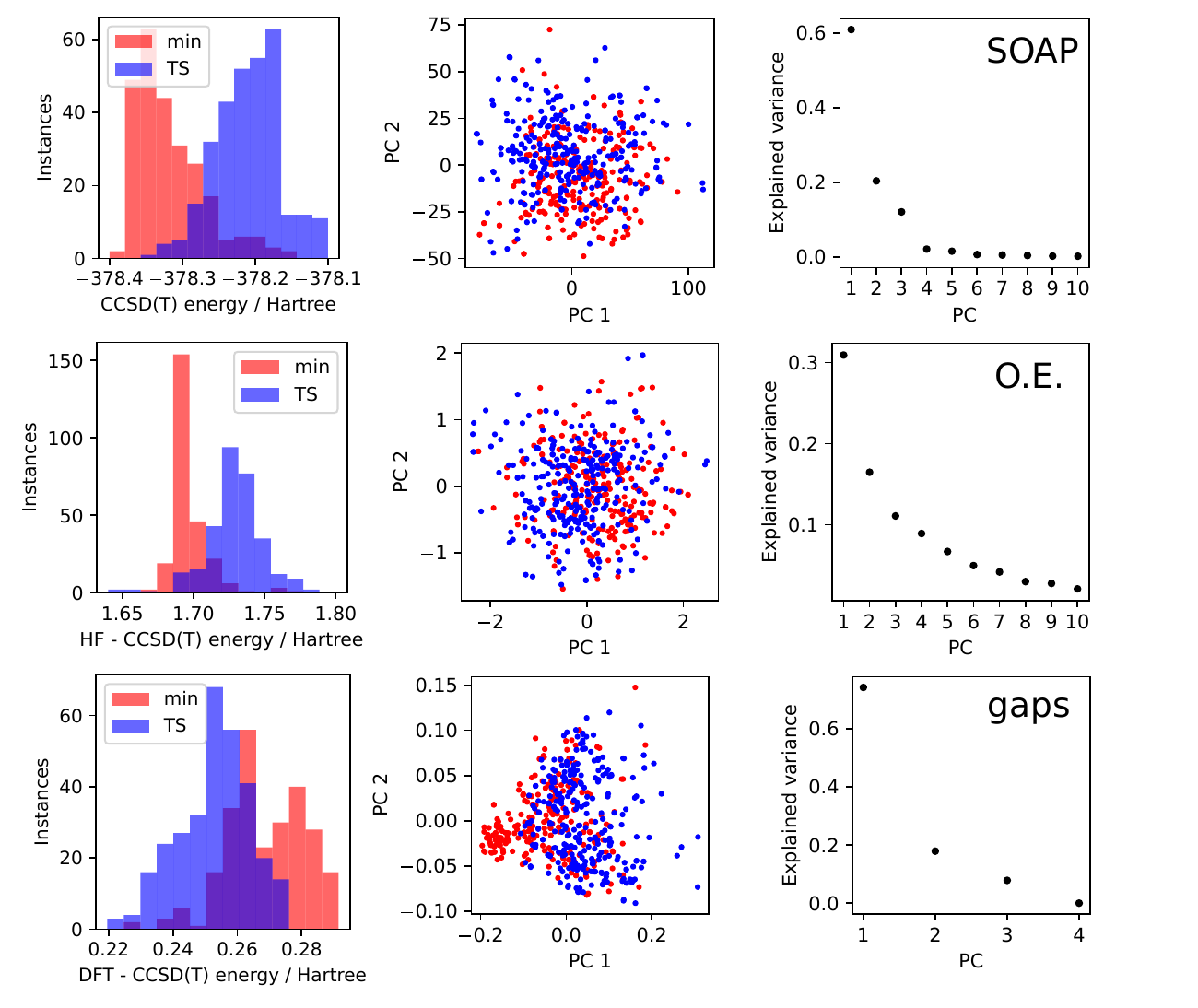}
    \caption{
    Energy analysis for the $N$-acetylimidazole CRN we explored.
    Left: histograms of different types of energies,
    where the minimum structure energies are given in red and the TS structure energies in blue.
    Top left: Histogram of the CCSD(T) total electronic energies.
    Middle left: Histogram of the difference between HF and CCSD(T) energies (i.e., correlation energies).
    Bottom left: Histogram of the difference between DFT and CCSD(T) energies.
    In the middle column,
    the PCA   
    of three different representations
    (SOAP, orbital energy (O.E.) spectrum, HOMO--LUMO gaps)    
    are shown for two principal components (PCs).
    In the right column, the fraction that explains the variance
    of the PCA is shown.
    }
    \label{fig:pca}
\end{figure}

We investigate now the $N$-acetylimidazole CRN we explored.
In the leftmost column of Figure \ref{fig:pca},
three histograms of energies are shown,
with the TS structures denoted in blue and the stable intermediates in red.
On top, the total CCSD(T) reference energies have two distinct peaks (0.15 Hartree apart)
with the TS structure energies being higher in energy.
In the middle panel, the histogram depicts the differences (i.e., the error) of HF energies to CCSD(T) energies (0.02 Hartree),
which corresponds to the electron correlation energy.
The correlation energy for the TS structures is higher.
In the bottom panel, the differences between DFT(PBE) and CCSD(T) is shown,
where the peak structure is not obvious because DFT somehow captures electron correlation in an approximate way and the deviation of DFT results from CCSD(T) ones is thus not the correlation energy.
Surprisingly, the DFT--CCSD(T) difference is smaller for TS energies than for minimum energies.

If we want to employ a particular molecular-similarity descriptor to predict energies, this descriptor should
encode information about whether a given structure is a TS structure or a minimum; that is., it should distinguish between those peaks.
We carried out a principal component analysis (PCA)
on the molecular structures in our $N$-acetylimidazole CRN 
in three representations. The results are shown in scatter plots of the first two principal components (PCs) in the middle column of Figure \ref{fig:pca}).
The input data is encoded as SOAP (top), orbital energies, the single-particle spectrum from an HF calculation, middle),
and four highest-occupied molecular orbital to lowest-unoccupied molecular orbital (HOMO--LUMO) gaps, the differences between each of HOMO-1 and HOMO, and the LUMO and LUMO+1 (bottom).
Apart from the cluster of minima on the left-hand side in the bottom panel, the features of the TS structures (blue) and of the local minimum structures (red) are not readily separable.
This indicates that they do not differ in the input dimension.

On the right-hand side of Figure \ref{fig:pca}, we show the decay of the eigenvalues of the PCA for each encoding. In SOAP,
the first two PCs explain around 81~\% of the variance.
In the second row, where the single-particle spectrum is shown,
the first two PCs explain about 48~\% of the variance.
Lastly, in the PCA for the HOMO--LUMO gaps,
the first two PCs explain around 92~\% of the variance.
The explained variance in this last example is naturally higher,
because the data is only encoded in four dimensions.
The visible separation stems mostly from the HOMO--LUMO gap,
which is in agreement with the expectation that the frontier orbitals carry much of the electronic information.
For the third PCA, we show the corresponding biplot in the Supporting Information.
It shows that the horizontal axis corresponds approximately to the energy difference 
from the HOMO to the LUMO and LUMO+1,
making them the important features in separating TS structures from minima.
    
It is in accordance with the elementary-step similarity dilemma that SOAP,
a structural descriptor,
is not able to separate TS structure from minima in the low-data regime,
as we would expect the largest discrepancy to be of electronic nature.
Therefore it is more surprising, that the electronic features,
are not able to separate them.
It does not bode well for the predictability of TS energies
if the corresponding structures are not (for the orbital energies) or only barely (for the HOMO-LUMO gaps) separable from the minima.

\begin{figure}
    \centering
    \includegraphics[width=0.5\textwidth]{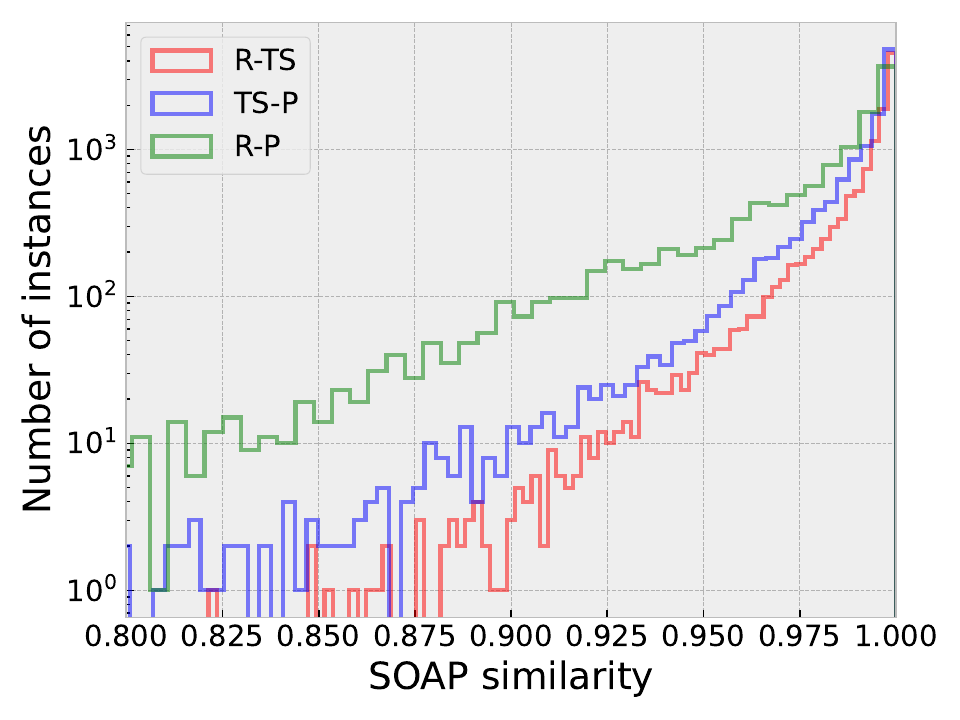}
    \caption{
    A histogram of the distances between
    reactant and TS structure (R-TS, red),
    TS structure and product (TS-P, blue),
    and reactant and product (R-P, green)
    of all 12,001 reactions in Grambow's data set.\cite{grambow2020}
    A flat distribution means overall reduced similarity.
    }
    \label{fig:soap_hist}
\end{figure}

Another way to look at this
is to compare the pairwise distances in a given descriptor space between
reactants and TS,
TS and product,
and reactant and product.
To demonstrate this effect and how it also holds for a more extensive data regime,
we will intermittently look at Grambows's data.\cite{grambow2020}
It contains
a total of 12,001 reactions from which
6388 (53.23~\%) had the greatest SOAP similarity for reactant and TS structure,
5074 (42.28~\%) for TS structure and product, and only
539 (4.49~\%) for the reactant to the product.
We can see the following in the histogram plot
(just the outlines of the buckets are shown for clarity)
in Figure \ref{fig:soap_hist} (left) over all pairwise SOAP similarities
(distances between SOAP descriptors):
the similarities between the reactants and products (green line, labeled R-P)
is much flatter,
i.e. number of instances is more distributed over a variety of similarities,
indicating reactant and product are in fact
not as similar as reactant and TS structure or TS structure and product:
The red line, denoting the distances between reactants and TS structures (R-TS),
and the blue line, denoting the distances between TS structures and products (TS-P),
are much more concentrated towards the right, indicating that more of the pairs are similar to each other.
On the very top right, the last bucket shows that red and blue have around 1000 instances more in the highest similarity class.

This is counter intuitive to
the electronic picture which lets us expect the TS to be more distinct
due to a potential multi-reference character, as we can see 
in SI Figure 9 (left), where we plotted the T1 diagnostic
(CCSD(T)-F12/cc-pVTZ-F12).
The T1 diagnostic is a measure of the single-reference character of a wavefunction, 
with values below 0.02 considered acceptable.\cite{lee1989}
The D1 diagnostic (SI Figure 9, right) is an additional assessment, with a common threshold of 0.05.\cite{lee2003}
Values above these thresholds may indicate multi-reference character.
The orange line shows that the TS structures generally show more multi-reference
character than the reactant and the product.
D1 and T1 diagnostic correlate well (SI Figure 10). 
A similar plot for the CM eigenvalues can be found in the Supporting Information.

\subsection{$\Delta$-machine learning vs. direct learning}
\label{sec:delta}

We compared two settings in our $\Delta$-ML approach.
In the first one, the target quantity is the difference between HF and CCSD(T) energies, i.e.,
the electron correlation energy.
In the second one, the difference of DFT(PBE) and CCSD(T) energies was studied.

\begin{figure}
\centering
\begin{minipage}{.5\textwidth}
  \centering
    \includegraphics[width=1\textwidth]{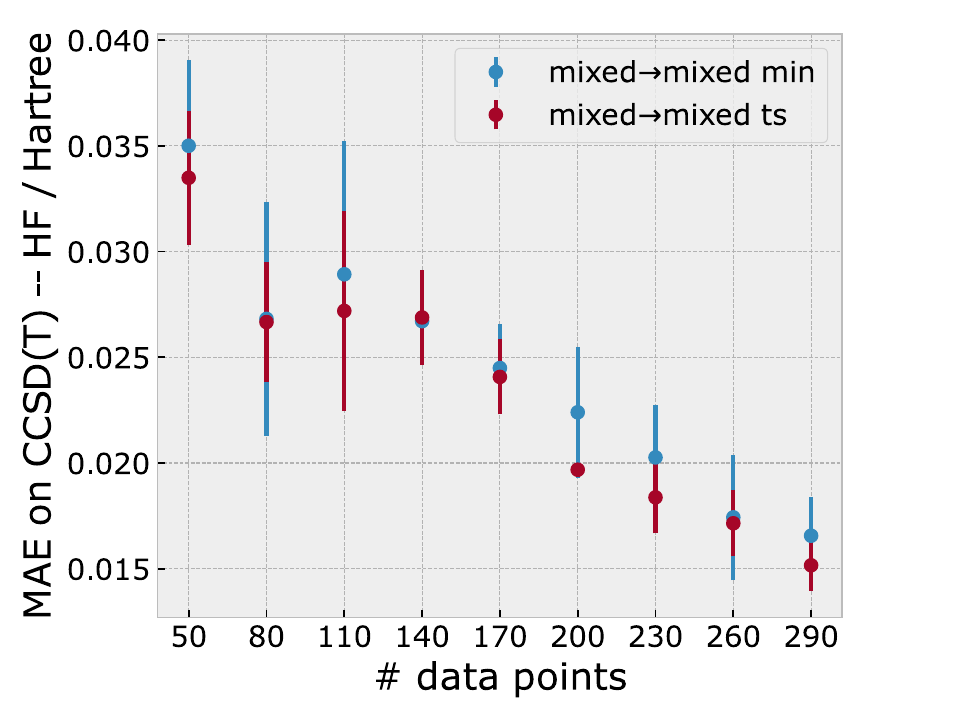}
\end{minipage}%
\begin{minipage}{.5\textwidth}
  \centering
    \includegraphics[width=1\textwidth]{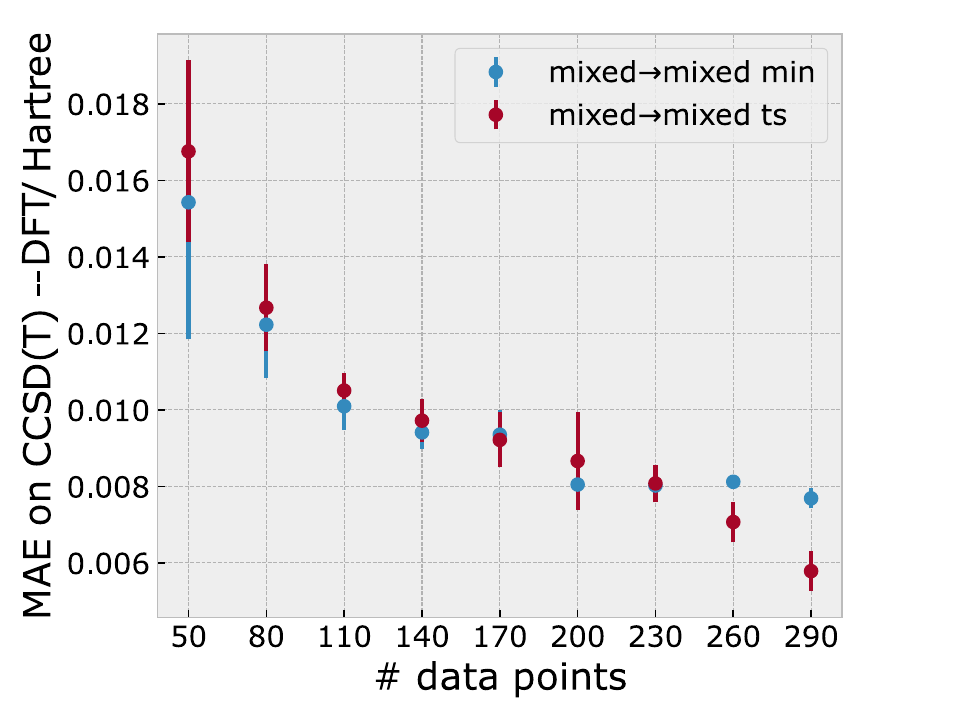}
\end{minipage}
\caption{
        Left: learning curve for the differences of CCSD(T) and HF energies.
        Right: learning curve for the differences of CCSD(T) and DFT(PBE) energies.
        The error is a mean absolute error (MAE) in Hartree 
        and the representation is SOAP.
        The error bars were obtained with 5 random initial draws.
        The algorithm was trained on a training set with both minimum and TS structures
        to predict a mixed set (`mixed'), where the error for each sub-population of minimum structures `mixed min' (blue) and TS structures `mixed ts' (red)
        is separated into two curves.
    }
    \label{fig:delta-hf-dft-cc}
\end{figure}

In Figure \ref{fig:delta-hf-dft-cc}, we compare these two settings.
On the left hand side of that Figure, we show the learning curve where the target is the difference between the CCSD(T) energy and the HF energy.
On the right, the learning curve is displayed for the difference between the CCSD(T) energy and the DFT(PBE) energy.
The error measure is the mean absolute error (MAE),
which is defined as the absolute difference between the estimate
and the reference.
The labels `mixed', `mixed min', and `mixed ts'
denote that we learn on a mixed training set composed of minima and TS structures.
From this, a test set of 50 minima (blue curve) and 50 TS structure energies (red line) are predicted.
We split the prediction set in order to show that the algorithm does not have more difficulty learning the TS structure energies
compared to the minimum energies.
We draw 50 instances at random 5 times to obtain the error bars.

Predicting the DFT to CCSD(T) $\Delta$ is considerably more accurate
(MAE at the end of 0.007 Hartree)
than the $\Delta$ of HF to CCSD(T)
(MAE of 0.016 Hartree).
This is the basic hypothesis behind $\Delta$ learning that a smaller discrepancy (see Figure \ref{fig:pca}) will lead to a smaller error and the inherent relationship between HF and CC does not improve learnability.
Both the minima and TS structures are predicted with similar accuracy,
which is consistent with the fact that the input space for minima and TS structures looked very similar (Figure \ref{fig:pca}).
From the histogram in Figure \ref{fig:pca},
we gathered that the peak distance for the
HF-CCSD(T) differences was 0.04 Hartree
and for the 
DFT-CCSD(T) differences was 0.02 Hartree.
If the GP regression as a first approximation predicts the mean 
(i.e. in the middle between the peaks),
it would incur an error for the HF-CCSD(T) difference of 0.02 Hartree,
and for the DFT-CCSD(T) difference of 0.01 Hartree (half the peak distance each).
This is consistent with the learning curves, 
because these errors are reached
at around 200 data points and 110 data points.

\begin{figure}
\centering
\begin{minipage}{0.5\textwidth}
  \centering
    \includegraphics[width=1\textwidth]{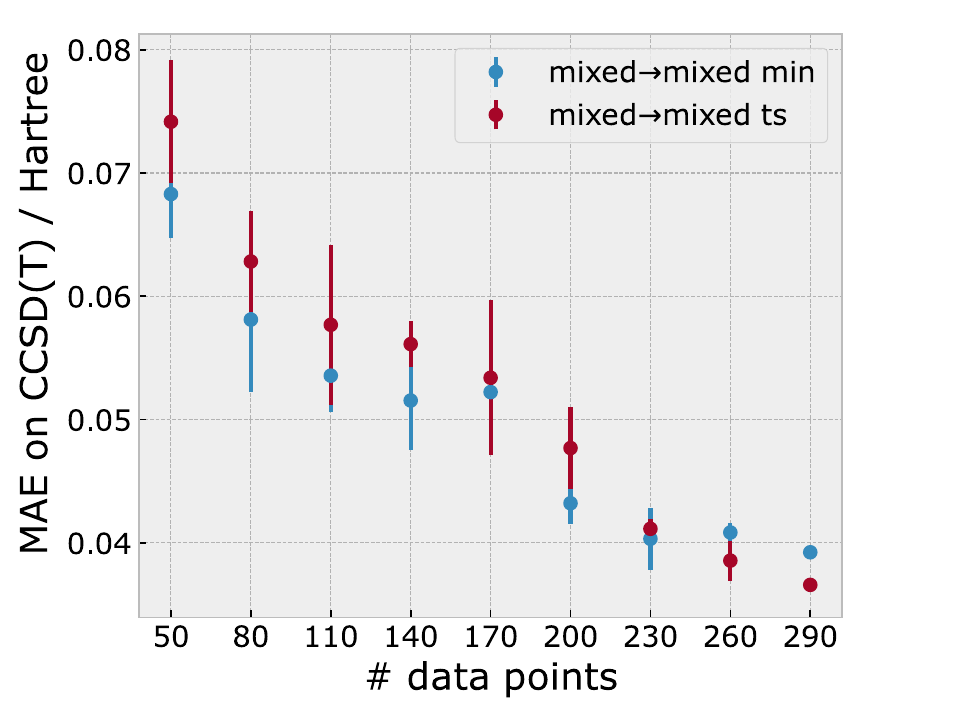}
\end{minipage}
\caption{
        Learning curve for learning CCSD(T) energies directly, avoiding $\Delta$-ML.
        The MAE is given in Hartree and the representation is SOAP.
    }
    \label{fig:cc-direct} 
\end{figure}

This result of the $\Delta$-learning can be compared with Figure \ref{fig:cc-direct},
where the CCSD(T) energies were learned directly,
which was overall worse
(minimal MAE of 0.04 Hartree at the last training step).
Again, compared to the peak distance of Figure \ref{fig:pca},
which was around 0.15 Hartree,
it is consistent with those findings,
as the learning curve starts at around 0.07 Hartree.
The relative learning curves can be found in the Supporting Information.

\subsection{Comparison of center-of-mass SOAP and variable-width SOAP}

In Ref. \citen{gugler2022}, we applied a SOAP formalism
where we compared the density of two molecules over the rotational integration
at the center of mass, as in Eq.~(\ref{eqn:integral}),
making the local descriptor a coarse global one.
This was chosen in order to introduce a translational invariance
that is less expensive than a pairwise approach\cite{bartok2013,de2016}
where every environment around an atom in one molecule is compared
to every environment around the atoms of the second molecule (see Section \ref{sec:soap-ext}).

\begin{figure}
\centering
\begin{minipage}{.5\textwidth}
  \centering
    \includegraphics[width=1\textwidth]{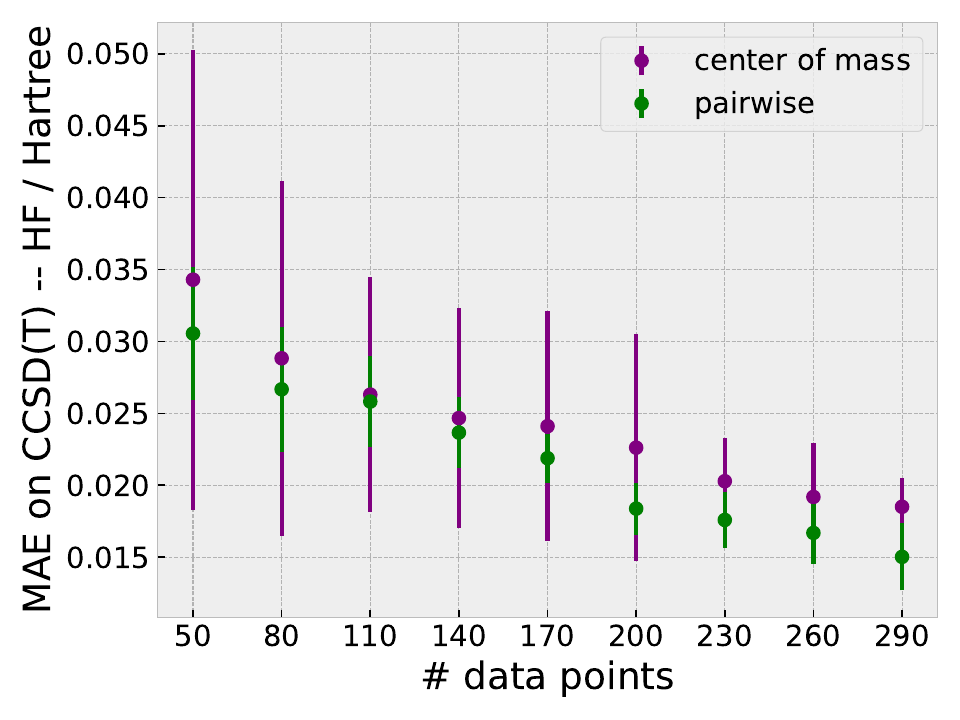}
\end{minipage}%
\begin{minipage}{.5\textwidth}
  \centering
    \includegraphics[width=1\textwidth]{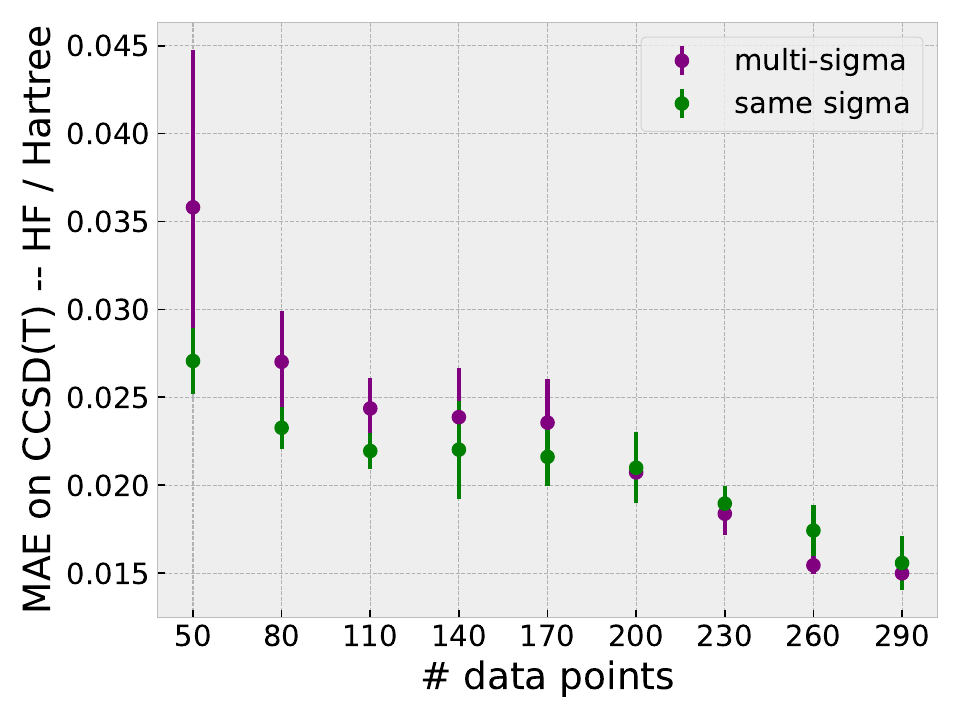}
\end{minipage}
\caption{
    Left:
    Comparison between learning curves for a GP regression on the differences between
    HF and CCSD(T) energies.
    The two descriptors are the center-of-mass SOAP definition\cite{gugler2022} (purple) and the standard SOAP definition\cite{bartok2013},
    that applies a pairwise formalism (green).
    Right:
    Comparison of a SOAP version with the width parameter $\sigma$
    kept constant for all element types (green)
    and a SOAP version where a different width parameter
    $\sigma_I$ is used
    for each element $I$ (purple).
    }
    \label{fig:com-multi-sigma}
\end{figure}

In Figure \ref{fig:com-multi-sigma} (left),
we show two learning curves to show the difference between the two options:
One for the SOAP encoding with the center-of-mass definition (purple)
and one with the pairwise definition (green).
Applying GP regression on the differences between the HF and CCSD(T) energies
shows that the learning curves do not differ significantly from each other.
Even though the pairwise SOAP appears to be better on average, it is all within the standard deviation.
We conclude that for the setting of small molecules in the low data regime,
the center-of-mass definition is sufficient.
As the trend in Figure \ref{fig:com-multi-sigma} (left) indicated, for larger
data regimes, it might be advantageous to employ the pairwise definition.
We also note that the aggregation scheme, Eq. (\ref{eqn:average}),
plays a role and more elaborate schemes may change this.\cite{de2016}

Another variation of SOAP uses different width parameters, $\alpha_I = 1/(2\sigma_I^2)$ (see Eq. \ref{eqn:soap})
for each element type.
When comparing standard SOAP with the same width parameter for every element (green)
and one with different widths (purple) depending on the van der Waals radius,
Figure \ref{fig:com-multi-sigma} (right),
we see that they are not significantly different.
Especially for a very small data set, it might be the case that the increased complexity of the descriptor
hampers the learning, whereas there is an insignificant difference for most of the rest of the learning process.
This is consistent with previous findings\cite{deringer2018} where the width parameter was optimized
as a hyper-parameter and ended up being the same for each element type.

\subsection{Learning local-minimum and TS structures from minima}
\label{sec:min-ts}

\begin{figure}
\centering
\begin{minipage}{.5\textwidth}
  \centering
    \includegraphics[width=1\textwidth]{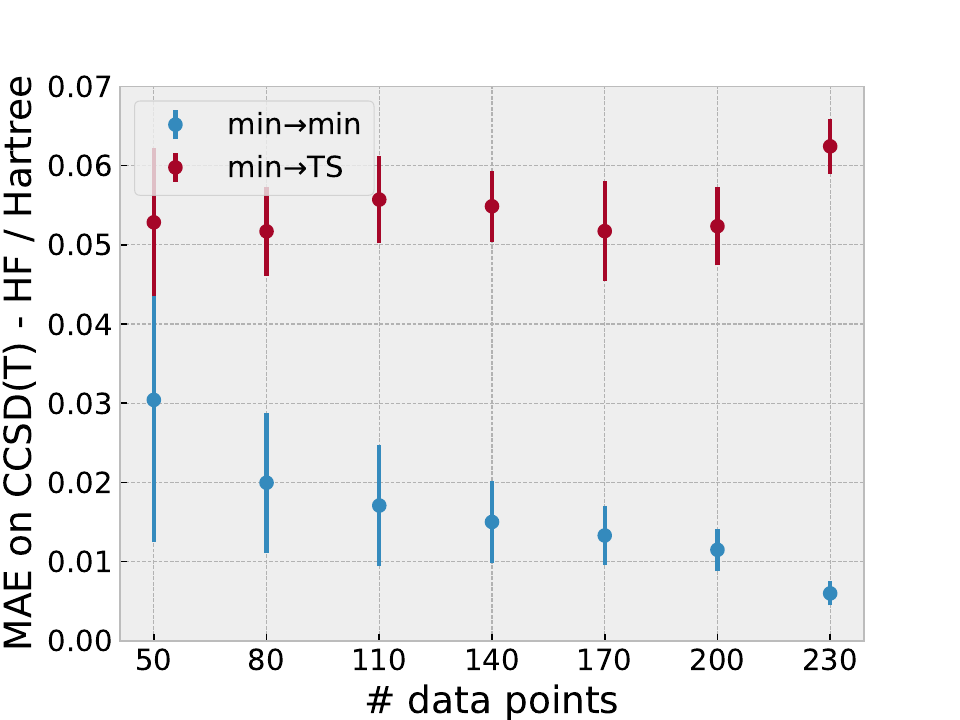}
\end{minipage}%
\begin{minipage}{.5\textwidth}
  \centering
    \includegraphics[width=1\textwidth]{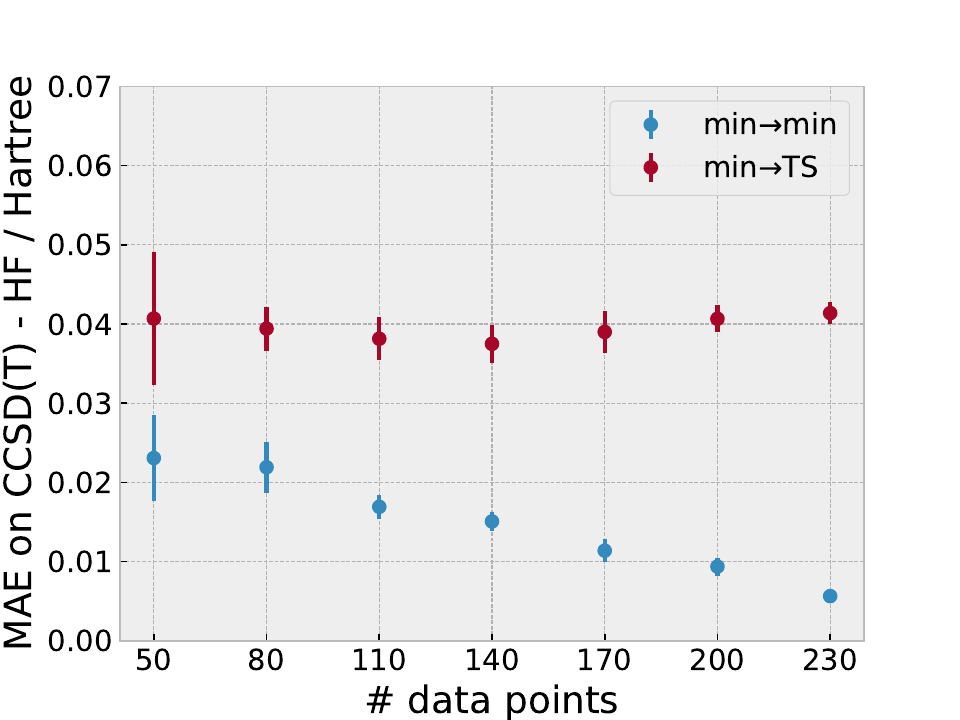}
\end{minipage}
\caption{
        Learning curve of GP regression
        with the eigenvalues of the Coulomb matrix (CM, left panel)
        and the Coulomb list (CL, right panel) descriptor.
        The target quantity is the difference between the HF energy and CCSD(T) energy
        and the error is given as an MAE.
        The blue markers denote the setting where the training set
        and the test set
        are comprised only of minima.
        The red markers denote the setting where we trained on minima but
        predict only TS structures.
        The error bars were obtained by randomly sampling the initial learning set of 50 data points 5 times.
    }
    \label{fig:learning-curves}
\end{figure}

We will now investigate how transferable a trained GP regression is when
trained on the correlation energy of structural minima (reactant and products)
and predicting the correlation energy of TS structures of our $N$-acetyl\-imidazole CRN.
As we can see in Figure \ref{fig:learning-curves}
(left panel for the CM eigenvalues, right panel for CL),
the MAE of the correlation energy decreases to less than 0.01~Hartree
for the case where the GP regression is trained on an increasing number of minima
and predicts correlation energies of the test set minima.
Training the GP regression on the correlation energy of minimum structures
but predicting the correlation energies of the TS structures (red) is unsuccessful.
The results do not converge and hover at around 0.05~Hartree (31.4~kcal/mol)
which indicates that the GP regression does not detect a pattern at all.
This can be explained from the PCA in Figure \ref{fig:pca}. 
Since the training only contains energies from the left peak of the minima
(left panel in Figure \ref{fig:pca})
and the input space does not separate the TS structures and the minima,
it seems that any TS structure will be estimated to be around the same energy as the minima. 
Turning it around and only learning on the TS structures and predicting the energies of the minima
results in a similar plot but with switched labels.
We furthermore confirm the hypothesis put forward in Ref. \citen{gugler2022} that the CM eigenvalues and
the CL are similarly suitable descriptors in terms of accuracy, since both of them compress much
information (i.e., $N^2$ interactions, where $N$ is the number of atoms) into a feature vector of size $N$.

\subsection{Combining electronic information with structural descriptors}

\begin{figure}
\centering
\begin{minipage}{.5\textwidth}
  \centering
    \includegraphics[width=1\textwidth]{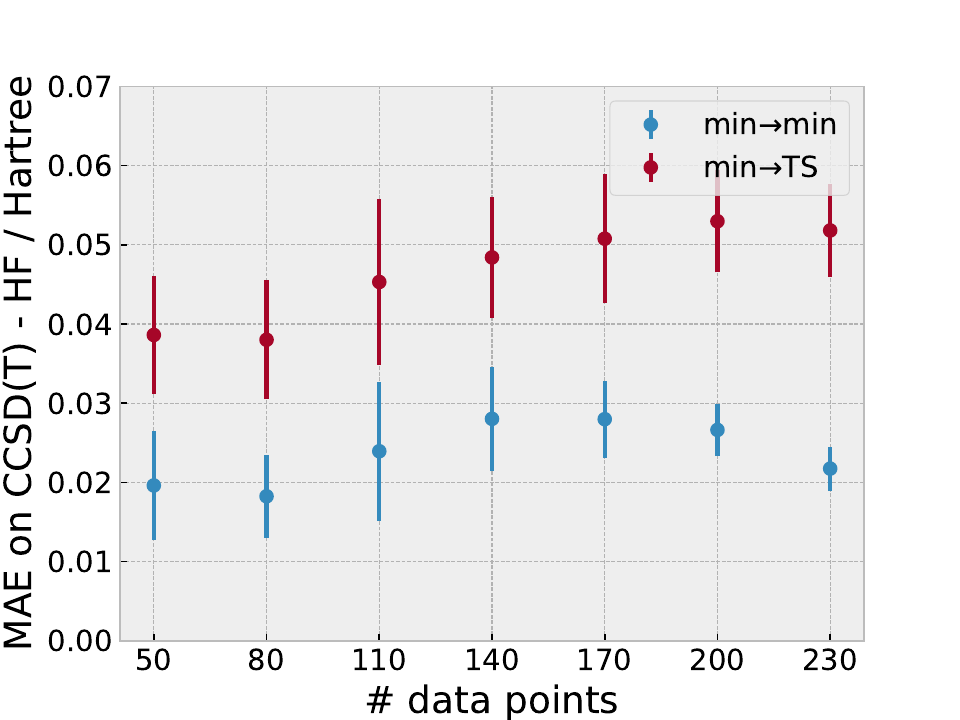}
\end{minipage}%
\begin{minipage}{.5\textwidth}
  \centering
    \includegraphics[width=1\textwidth]{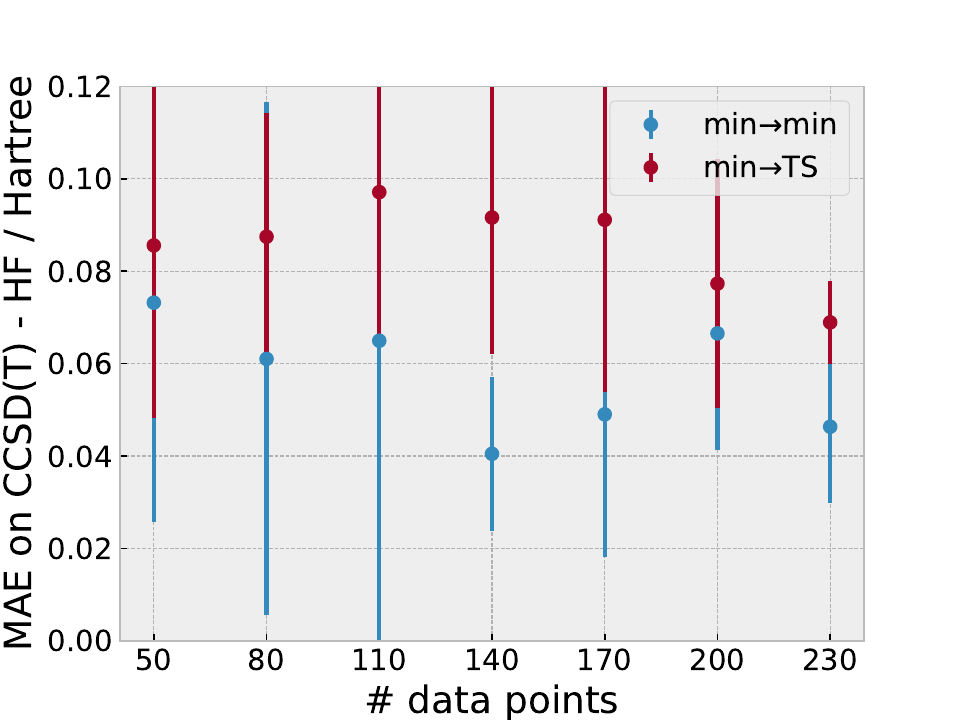}
\end{minipage}
\caption{
    Learning curve of GP regression with electronic information.
    Left: full single particle spectrum, 
    right: gaps of all HOMO--LUMO gaps between the HOMO, HOMO-1, LUMO, and LUMO+1 energy levels.
    The blue dots denote the training scenario where the training set and the test set are comprised of only minima,
    whereas the red dots denote the scenario where the GP is trained on minima but predicts TS structures.
    The error is measured as MAE of the approximate correlation energy, which is the difference
    between the CCSD(T) energy and the HF energy, measured in Hartree.
    The error bars were obtained by randomly sampling the initial learning set of 50 data points 5 times.
    }
    \label{fig:learning-curves_elec}
\end{figure}

In Figure \ref{fig:learning-curves_elec},
we employed the single-particle spectrum (left panel)
and the HOMO--LUMO gaps between the each of the HOMO and HOMO-1, and the LUMO and LUMO+1 energy levels
of a HF calculation
as descriptors.
As in Section \ref{sec:min-ts},
we trained the GP regression on a random set of 50 initial structures.
This random selection was carried out 5 times to obtain error bars.
The training set contained only minimum structures.
Both the learning curve for predicting other minimum energies (blue) as well as the one for the energies of TS structures (red)
did not converge, which
implies that the GP regression did not learn a pattern.
This is in accordance with the evidence from 
the PCA in Figure \ref{fig:pca}, 
indicating that these simple electronic descriptors are not separating the TS structures well from the minima.
Especially for the gaps encoding, the prediction appears to be dependent on the initial conditions, 
indicated by the large variance in each prediction.
Unfortunately, this shows that much more elaborate procedures need to be devised to capture the intricacy of the
the TS structure energies.

\subsection{General Discussion}

Our findings highlight the challenges of predicting TS energies within low-data-regime CRNs based on molecular similarity descriptors that primarily encode structural information.
Although these descriptors capture similarities in molecular structures, they fall short in distinguishing TS structures from connected minimum structures in terms of differences in their electronic structures.
This limitation becomes apparent when attempting to predict TS energies with models only trained on data obtained for local minimum structures. The descriptors cannot adequately represent the intricate differences in electronic structures, at least not in the low-date regime.

In the context of CRNs, where reactants, products, and TS structures are chemically related, one might expect structural descriptors to suffice for energy predictions.
However, our results demonstrate that structural similarity might not be high enough in a CRN and that structural similarity does not necessarily translate into electronic similarity, especially for TS structures which often exhibit multi-configurational character due to stretched bonds or transient electronic states.
This discrepancy underscores the elementary-step similarity dilemma we observed, where TS structures are structurally more similar to the connected minimum structures, but electronically distinct from them.

Comparing our approach to existing methods in the literature, we note that many studies focus on predicting activation barriers directly by modeling reactions as integrated entities, often combining reactant and product information into a single descriptor.\cite{choi2018,vangerwen2024}
Such approaches effectively provide context for each specific barrier, making it easier to capture the relevant electronic changes during the reaction.
In these cases, the explicit modeling of TS structures is also avoided.
By contrast, our approach aims to predict absolute and, more successfully, relative energies of individual molecular structures within a single CRN, rather than focusing solely on activation barriers or on broad sections of chemical reaction space.
This approachs offers the advantage of reusability of energy predictions; that is, for different reactions involving the same species, an energy can be reused when calculating an energy barrier.
But it also presents additional challenges: Predicting TS energies requires accurately capturing subtle electronic effects that are not easily encoded by structural descriptors alone.
Our results suggest that simple electronic features, such as HOMO--LUMO gaps or orbital energies, are insufficient to bridge this gap.
In this context, T1 or D1 diagnostics may offer a complementary approach to ML for TS energies beyond standard structural or orbital-based descriptors.

\section{Conclusions}
\label{sec:conclusion}

For analyzing chemical reaction networks, electronic energies of
stable intermediates and transition state structures are key input and need 
to be calculated with high accuracy.
However, since reaction networks are vast, their exploration requires 
fast electronic structure models that trade speed for accuracy.
Accurate reference calculations will only be available for some
of the nodes in such a network owing to their computational costs.
The results of these reference calculations can then be transferred to
related (similar) structures based on Bayesian uncertainty quantification.\cite{simm2018} 
The context provided by a reaction network guarantees chemical relatedness
as each structure is only an elementary step away from another structure,
which makes kernel-based machine learning (such as Gaussian process regression) especially suitable for this problem.

In this work, we studied machine learning to predict total electronic energies as well as $\Delta$-machine learning corrections for the correlation energy.
While it is difficult to predict transition state energies
from learning on stable intermediates only,
predicting minima shows characteristic learning convergence.

We analyzed the role of molecular similarity descriptors, that can be related to electronic structure theory. \cite{gugler2022}
and the
inclusion of basic electronic structure information.
Our analysis showed that structural descriptors like SOAP and the CM eigenvalues can capture the similarity between minimum structures, but they struggle to distinguish TS structures due to the distinct electronic features of the latter. ML models trained on minima were unable to accurately predict TS energies, highlighting the limitations of relying solely on structural similarity in this context.
We also observed that over a large reaction data set\cite{grambow2020}.

We demonstrated that $\Delta$-ML approaches, which learn the difference between a low-level and a high-level electronic structure method, can improve prediction accuracy over direct learning of total energies. 
Learning the difference between DFT energies and CC energies showed better performance than learning the electron correlation energy directly from HF energies. This suggests that using a more accurate base method in $\Delta$-ML can enhance the model's predictive capabilities.

Additionally, we explored the incorporation of electronic structure information, such as orbital energies and HOMO--LUMO gaps, into the ML models. However, these electronic descriptors did not significantly improve the prediction of TS energies when combined with structural descriptors. This finding indicates that more sophisticated descriptors or alternative approaches are necessary to capture the complex electronic behavior of TS structures.

We also compared SOAP evaluated locally at the center of mass as in Ref. \citen{gugler2022} to SOAP evaluated for each pairwise fashion and then averaged.
For the small molecules studied in this CRN, the pairwise SOAP descriptor did not afford better learning.
Furthermore, we studied another SOAP variation that has a variable exponent $\alpha_I$ which did not improve accuracy. This might be taken as further evidence that SOAP is a suitable approximation to the electron density.

Overall, our work underscores the importance of developing descriptors that can capture both structural and electronic features of molecular structures, paving the way for more accurate and efficient predictions of various energies in chemical reaction networks.

\providecommand{\refin}[1]{\\ \textbf{Referenced in:} #1}

\end{document}